# All-optical graph representation learning using integrated diffractive photonic computing units

Tao Yan[1,4,*], Rui Yang[1,2,3,*], Ziyang Zheng[1,2,3], Xing Lin[2,4,5,6,§], Hongkai Xiong[3,§], Qionghai Dai[1,4,5,6,§]

[1]Department of Automation, Tsinghua University, Beijing 100084, China
[2]Department of Electronic Engineering, Tsinghua University, Beijing 100084, China
[3]Department of Electronic Engineering, Shanghai Jiao Tong University, Shanghai 200240, China
[4]Institute for Brain and Cognitive Sciences, Tsinghua University, Beijing 100084, China
[5]Beijing National Research Center for Information Science and Technology, Tsinghua University, Beijing 100084, China
[6]Beijing Laboratory of Brain and Cognitive Intelligence, Beijing Municipal Education Commission, Beijing 100084, China

§Correspondence to: lin-x@tsinghua.edu.cn, xionghongkai@sjtu.edu.cn, daiqh@tsinghua.edu.cn
*These authors contributed equally.

## Abstract

Photonic neural networks perform brain-inspired computations using photons instead of electrons that can achieve substantially improved computing performance. However, existing architectures can only handle data with regular structures, e.g., images or videos, but fail to generalize to graph-structured data beyond Euclidean space, e.g., social networks or document co-citation networks. Here, we propose an all-optical graph representation learning architecture, termed diffractive graph neural network (DGNN), based on the integrated diffractive photonic computing units (DPUs) to address this limitation. Specifically, DGNN optically encodes node attributes into strip optical waveguides, which are transformed by DPUs and aggregated by on-chip optical couplers to extract their feature representations. Each DPU comprises successive passive layers of metalines to modulate the electromagnetic optical field via diffraction, where the metaline structures are learnable parameters sharing across graph nodes. DGNN captures complex dependencies among the node's neighborhoods and eliminates the nonlinear transition functions during the light-speed optical message passing over graph structures. We demonstrate the use of DGNN extracted features for node and graph-level classification tasks with benchmark databases and achieve superior performance. Our work opens up a new direction for designing application-specific integrated photonic circuits for high-efficiency processing large-scale graph data structures using deep learning.

## Introduction

Deep learning technologies[1] have achieved enormous advances in a wide range of artificial intelligence (AI) applications, including computer vision[2], speech recognition[3], natural language processing[4], autonomous vehicles[5], biomedical science[6], etc. The core is to leverage multi-layer neural networks to learn hierarchical and complicated abstracts from big data, driven by the continuous development of integrated electronic computing platforms, such as CPUs[7], GPUs[8], TPUs[9], and FPGAs[10]. However, the electronic computing performance is approaching its physical limit and faces large difficulties to keep pace with the increase in demand of AI development, which is a common plight in a broad range of applications requiring large-scale deep neural models. In recent years, there has been growing research of interest in photonic computing to use photons as the computing medium to construct photonic neural networks by utilizing its advanced properties of high parallelism, minimal power consumption, and light-speed signal processing.

Numerous photonic neural network architectures have been proposed to facilitate complex neuro-inspired computations[11-13], such as diffractive neural networks[14-21], optical interference neural networks[22-24], photonic spiking neural networks[25-27], and photonic reservoir computing[28-30]. Existing architectures have been most successful in processing data with regular structures in the form of vectors or grid-like images. Nevertheless, various scientific fields analyze data beyond such underlying Euclidean domain. As typical representatives, graph-structured data, which encode rich relationships (i.e., edges) between entities (i.e., nodes) within complex systems, are ubiquitous in the real world, ranging from chemical molecules[31] to brain networks[32]. To process the graph-structured data, graph neural networks (GNNs)[33-39] have been developed as a broad new class of approaches that are able to integrate local node features and graph topology for representation learning. Among these models, message passing-based GNNs have major advantages of flexibility and efficiency by generating neural messages at graph nodes and passing along edges to their neighbors for feature updates. It has been successfully applied in many graph-based applications, including molecule property prediction[31], drug discovery[40], skeleton-based human action recognition[41], spatio-temporal forecasting[42], etc. However, how to effectively take advantage of photonic computing to benefit graph-based deep learning still remains largely unexplored.

In this article, we propose the diffractive graph neural network (DGNN), a novel photonic GNN architecture that can perform optical message passing over graph-structured data. DGNN is built upon the foundation of integrated diffractive photonic computing units (DPUs) for generating the optical node features. Each DPU comprises the successive diffractive layers implemented with metalines[43-45] to transform the node attributes into optical neural messages, where the strip optical waveguides are deployed to encode the input node attributes and output the transformed results. The optical neural messages sent from node neighborhoods are aggregated by employing optical couplers. In DGNN architecture (Fig. 1), the DPUs can be cascaded horizontally to enlarge the receptive field to capture complex dependencies from the arbitrary size of neighboring nodes. Besides, the DPUs can also be stacked vertically to extract higher-dimensional optical node features for increasing its learning capacity, inspired by the multi-head strategy used in numerous modern deep learning models, e.g., Transformer[46] and graph attention networks[37]. Based on this scalable optical message-passing scheme, we first demonstrate the semi-supervised node classification task, where the DGNN extracted optical node features are fed into an optical or electronic output classifier to determine the node category. The results show that our optical DGNN achieves competitive and even superior classification performance with respect to the electronic GNNs on both synthetic graph models and three real-world graph benchmark datasets, i.e., two citation networks and one Amazon Co-purchase graph. Furthermore, DGNN also supports graph-level classification, where the additional DPUs are utilized to aggregate all-optical node features into a graph-level representation for classification. The results on skeleton-based human action recognition demonstrate the effectiveness of our architecture for the task of graph classification.

## Results

**General GNN design principle.** A graph structure of $N$ nodes is represented as a tuple $\mathcal{G} = (\mathcal{V}, \mathcal{E}, \mathbf{A})$, where $\mathcal{V} = \{v_i\}_{i=1}^N$ is the node-set with $i$ denoting the node indices, $\mathcal{E} \subseteq \mathcal{V} \times \mathcal{V}$ is the edge-set, and $\mathbf{A} \in \mathbb{R}^{N \times N}$ is the adjacency matrix encoding the connection between graph nodes. Fig. 1a shows a toy example of a graph with six nodes, where each node $v_i \in \mathcal{V}$ is attached with a three-dimensional attribute $\mathbf{x}_i$. One prominent and powerful approach in most existing GNNs to learn effective node representations is the message passing scheme[34-38], as depicted in Fig. 1b. Each node aggregates neural messages sent from local neighborhoods during each iteration $l$ ($l = 1, \ldots, L$) of the message passing procedure:

$$\mathbf{m}_j^{(l)} = \text{MSG}^{(l)}\left(\mathbf{h}_j^{(l-1)}\right), \tag{1}$$

$$\mathbf{h}_i^{(l)} = \text{AGG}^{(l)}\left(\mathbf{h}_i^{(l-1)}, \left\{\mathbf{m}_j^{(l)} | j \in \mathcal{N}(i)\right\}\right), \tag{2}$$

where $\mathbf{m}_j^{(l)}$ is the neural message of $j$-th node, $\mathcal{N}(i)$ is the neighboring node indices of node $v_i$, $\mathbf{h}_i^{(l)}$ is the updated features of node $v_i$ after $l$ iterations of message passing, $\mathbf{h}_i^{(0)} = \mathbf{x}_i$ is the initial node attributes, $\text{MSG}^{(l)}(\cdot)$ is a neural network shared across graph nodes to perform the feature transformation and generate neural messages, and $\text{AGG}^{(l)}(\cdot)$ is a function that aggregates messages sent from local neighborhoods. To generate the graph-level representation $\mathbf{h}_\mathcal{G}$, a read-out function $\text{Read-out}(\cdot)$ can be applied to aggregate all node features into a vector after $L$ rounds of message passing:

$$\mathbf{h}_\mathcal{G} = \text{Read-out}\left(\mathbf{h}_i^{(L)} | v_i \in \mathcal{V}\right). \tag{3}$$

With the extracted node/graph-level features, we can perform node/graph-level classification task by feeding the features to the output classifier, and jointly learn model parameters via the end-to-end error backpropagation training method. In the following, we elaborate the design of DGNN to implement these critical operations by using on-chip optical devices and modules.

**DGNN architecture for optical message passing.** The DGNN architecture is illustrated in Fig. 1c-d, which comprises all-optical devices and modules to implement the $\text{MSG}(\cdot)$, $\text{AGG}(\cdot)$, and $\text{Read-out}(\cdot)$ functions in Eqs. (1-3). Specifically, the input node attributes are encoded by modulating the amplitude or phase of the coherent light, which can be realized via on-chip optical modulators, e.g., Mach-Zehnder interferometers (MZIs)[22]. The input optical field of each node attribute passes through a single-mode waveguide with transverse-electric (TE) polarization and is coupled into the integrated DPU module. The DPU module achieves the $\text{MSG}(\cdot)$ function by using successive layers of 1D metalines as the diffractive layers to modulate the input optical field (see Methods and Supplementary Fig. S1-S5). The metaline is a 1D etched rectangle silica slot array in the silicon membrane of silicon-on-insulator (SOI) substrate that forms as diffractive meta-atoms. The modulation coefficients of a diffractive meta-atom in metaline are determined by the height and width of the slot (see Supplementary Fig. S1). The neural message of each node is generated by coupling the output optical field of DPU with the single-mode output waveguides, where the number of optical waveguides $m$ determines the message dimensionality. We set $m = 2$ to avoid the waveguide crossing during the aggregation of neural information from neighboring nodes in this work, which can be scaled to arbitrary size in principle for further increasing the learning capability (see Supplementary Fig. S6).

The $\text{AGG}(\cdot)$ function is realized by the optical Y-coupler (see Fig. 1c), where the feature aggregation of two-dimensional optical neural messages over two nodes wouldn't cause the waveguide crossing. Thus, the architecture can aggregate information from arbitrary size of neighborhood by stacking the building block of DPU horizontally. The only waveguide crossing happens during the injection of light from the coherent source to DPU modules (see Supplementary Fig. S7), which can be well-addressed by the existing waveguide crossing technology to minimize the signal crosstalk and energy loss[47]. To further enhance the expressive power of two-dimensional node features, $P$ independent heads can be vertically stacked in parallel as shown in Fig. 1d to produce $2P$-dimensional optical features for graph nodes following the multi-head strategy. Besides, the procedure of a single round optical message passing for feature updating can be further stacked to perform multiple rounds of message passing.

To enable graph-level learning, the read-out function $\text{Read-out}(\cdot)$ can be realized by applying additional DPUs to aggregate all multi-head optical node features into the optical graph features. First, each head of a graph node in Fig. 1d for producing optical node features is cascaded with a read-out DPU with two input waveguides and two output waveguides. Then, we aggregate the updated $2P$-dimensional optical node features of all nodes over each independent head by using the optical Y-coupler to perform a two-by-two optical aggregation, which obtains the $2P$-dimensional optical graph features. By feeding the extracted node/graph-level features into the output optical or electronic classifier, corresponding to the DGNN-O or DGNN-E, respectively, the node/graph classification tasks are performed. The modulation coefficients of all diffractive meta-atoms are jointly optimized via the end-to-end error backpropagation training method.

**Node classification using semi-supervised learning.** We apply the DGNN for semi-supervised node classification, which is one of the major AI tasks that GNNs have achieved significant success so far. Given the graph where each node is attached with vector-based attributes and a subset of graph node labels, the node classification task is to infer the labels for the remaining nodes. To scale-up GNNs for tackling large graph datasets, we adopt the PPRGo[38] model to directly capture the high-order neighborhood information with a single AGG(·) process (see Methods). This avoids the exponential neighborhood expansion problem during the multiple rounds of message passing and eliminates the nonlinear transition function. For each target node $v_i$, we use the DPU to implement $MSG(\mathbf{x}_i)$ and then aggregate optical features of nodes with the top-$k$ largest scores according to its personalized PageRank vector. After the training process of DGNN with the DPU settings detailed in Methods, the optical modulation coefficients are optimized and the slot width of diffractive meta-atoms in the metalines are determined. We validate the superior classification performance of the DGNN on both the synthetic graph data and three real-world large-scale graph datasets, i.e., Cora-ML[48,49], Citeseer[50], and Amazon Photo[51,52], using both photonic finite-difference time-domain (FDTD) and analytical model evaluation.

The *synthetic graph* in Fig. 2a is generated by using the stochastic block model (SBM)[53] to simplify the task and reduce the computational complexity for FDTD evaluation (see Methods). In this example, the DGNN is trained by configuring a single head, i.e., $P = 1$, that generates a two-dimensional neural message for each target node from a three-dimensional node attribute. The layout of the DPU module is shown in Fig. 2b(bottom), which includes the corresponding input and output waveguides and three layers of metalines. Each metaline has 90 diffractive meta-atoms, with each meta-atom size of 300 $nm$. We set the binary modulation for meta-atoms with every 3 consecutive elements to be the same, i.e., the same silica slot width and height, in order to consider the fabrication capability of existing silicon photonics foundry and reduce the modulation error of the analytical model with respect to FDTD (see Supplementary Fig. S2). The comparisons of output optical neural message of DPU module between the analytical model and FDTD are evaluated in Supplementary Fig. S3 and S4. Fig. 2b(top) shows the optical field propagation of the DPU module using FDTD evaluation for an exemplar node.

We use the DGNN-E for node classification of the synthetic graph that feeds the intensity detection of the calculated optical neural message to an electronic fully-connected layer, where the numbers of input nodes and output neurons are equivalent to the feature dimension and category number, respectively. We further update each node's representation by aggregating features with different $k$ to re-train the output electronic classifier for validating the effectiveness of the optical node representation. The classification results under different $k$ are shown in Fig. 2c. The DGNN-E evaluated with FDTD achieves comparable performances with respect to the analytical model with system errors included (see Methods), and both are superior to the electronic PPRGo GNNs and multi-layer perceptrons (MLP) under the same network size. The results demonstrate the distinguishable of the extracted optical node representations for the node classification. Furthermore, the output waveguide can be tapered to couple the larger regions of the output optical field for improving the DPU energy efficiency without decreasing the classification performance (see Supplementary Fig. S5).

On *real-world benchmark graph databases*, we construct the DGNN architecture by setting the top-$k$ node number for feature aggregation to 8 and the head number to 4, i.e., $k = 8, P = 4$. Thus, each node generates eight-dimensional optical features in total, two-dimension for each head, from the pre-processed node attributes (see Methods and Supplementary Table S1). For the DGNN-E with electronic output classifier, the intensity of optical features detected with photodetectors is fed into an electronic fully-connected layer, similar to the process on the synthetic graph. For the DGNN-O with optical output classifier, the 8 output waveguides of optical features are directly coupled with a classifier DPU module composed of 6 diffractive layers with other settings the same as the DPU modules for generating optical neural messages. The classification results are detected by the photodetectors, each corresponding to one category, where the category of input is determined by finding the target photodetector with the maximum detected optical signal[14].

We report in Table 1 the analytical test accuracies of semi-supervised node classification, i.e., the transductive learning, of DGNN on three benchmark graphs and compare them to electronic computing approaches of linear PCA, nonlinear MLPs, and nonlinear PPRGo GNNs, including PPRGo-S and PPRGo-WS (see Methods). Both the MLP and feature transformation of PPRGo are configured by using a fully-connected neural network with a hidden layer size of 8, equivalent to the node feature dimension. The test accuracy convergence plots of DGNN-O and DGNN-E are shown in Fig. 3a, b, respectively. Besides, we also evaluate the DGNN-E using binary diffractive modulations with system errors by including the different amounts of Gaussian noise in Supplementary Fig. S8. Although the convergences fluctuate due to the rounding operation, the re-training scheme can be adopted to achieve stability and obtain even higher accuracy by fixing the learned modulation coefficient and re-training the output classifier. The confusion matrices of test results using binary diffractive modulation with a system error, implemented by including the standard deviation of 0.3 of Gaussian noise, on three benchmark databases are shown in Fig. 3d-f, which achieves test accuracies of 86.7%, 74.4%, and 93.6% on Cora-ML, Citeseer, and Amazon Photo, respectively. Overall, the results in Table 1 reveal the following facts: (1) models that exploit the graph structure substantially outperform models that ignore the graph structure; (2) the all-optical inference of DGNN-O achieves competitive performance with the PPRGo; (3) DGNN-E achieves 1.6% higher classification accuracy than PPRGo-S on the Cora-ML database, showing the optical modules of DGNN for implementing MSG(·) and AGG(·) are even more effective than the electronic message passing; (4) As the feature aggregation is essentially a low-pass filter[54] that can suppress the noise to a certain extent, the binary diffractive modulation with system error included also achieves competitive performance on all the graph benchmarks, demonstrating the robustness of architecture to system noise.

Furthermore, we conduct ablation studies of DGNN-E architecture on the neighborhood size $k$ for feature aggregation and the number of heads $P$ for the output classifier (Supplementary Fig. S9). Note that $k = 1$ refers to the architecture without message passing, which degenerates to the plain diffractive neural network[14]. By increasing $k$ from 1 to 8, the test accuracy on Cora-ML monotonically increases and gains over 9% compared with the diffractive neural network (Fig. S9a), demonstrating the functionality of feature aggregation for node classification. Besides, the multi-head scheme significantly improves the test accuracy by generating a higher dimension of optical features (Fig. S9b). We also demonstrate that input node attributes can be flexibly encoded into the amplitude or phase of input optical fields with comparable model performance (Fig. S9c). And the importance of diffractive modulation for performing the feature transformation is shown in Fig. S9d and S9e. To visualize the generated optical feature representations for all graph nodes, we apply the t-SNE[55] on the detected intensity values of optical node features for DGNN-E with $k = 8$ and $P = 4$. As illustrated in Fig. 3c and Supplementary Fig. S10, the t-SNE plots show that the node features exhibit discernible clustering across different classes in the projected 2D space, verifying the effectiveness of the optical implementations of $\text{MSG}(\cdot)$ and $\text{AGG}(\cdot)$ functions in DGNN architecture. In addition to semi-supervised transductive learning, we also evaluate the inductive reasoning aptitude of DGNN (see Methods). The inductive node classification results on the three benchmarks are shown in Supplementary Fig. S11 and Table S2, where DGNN still outperforms or achieves competitive performance with all baselines.

**Graph classification for skeleton-based human action recognition.** We validate the performance of DGNN on graph-level classification by applying it for the task of skeleton-based human action recognition. The skeleton data are sequences of frames with each frame containing a set of 3D joint coordinates of the target recorded by sensors, with which the task is to predict the category of action in each sequence. In this work, we adopt the UTKinect-Action3D database[56] for evaluation, which contains the RGB, depth, and skeleton videos of 10 subjects performing each action two times that are captured by a single stationary Kinect V1 at a frame rate of 30 fps. Here, we select 6 out of 10 types of actions from the skeleton-based data, including walk, sit down, stand up, pick up, wave hands, and clap hands. The graph structure of a skeleton is shown in Fig. 4a, which contains the $(x, y, z)$ locations of 20 joints at each frame.

The workflow of DGNN architecture is illustrated in Fig. 4b. We implement the $\text{MSG}(\cdot)$ and $\text{AGG}(\cdot)$ functions using 4 heads that generate an 8-dimensional optical node feature for each joint, where the neural messages at each joint are aggregated from their direct neighbors. At each skeleton frame, the head is cascaded with a DPU to perform $\text{Read-out}(\cdot)$ and aggregates all node features into an 8-dimensional optical graph feature. Similar to the previous work[18], we divide each sequence with a length of $M$ into numbers of sub-sequences with the same length of $n$ ($n \ll M$). Then, $n$ graph-level representations are concatenated to be fed into the output classifier for action recognition of the sub-sequence. We set $n = 6$ in this study, resulting 48-dimensional optical node features for each sub-sequence, which are fed into an electronic fully-connected neural network layer to determine the sub-sequence category. The video category is obtained by applying the winner-takes-all strategy[18,30] on all video sub-sequences. To ensure the credibility of the evaluation, we perform 5-fold cross validation on the 20 subjects with 6 actions, i.e., 120 videos and 2512 sub-sequences, and report the average sub-sequence accuracy and video accuracy.

Our DGNN architecture achieves test sub-sequence accuracy of 83.3% and video accuracy of 90.0%, verifying the effectiveness of the proposed method on graph-level learning. We visualize the results of sub-sequence action recognition for the categories of the walk and wave hands, as shown in Fig. 4c and Supplementary Fig. S12, respectively. It's obvious that the optical $\text{MSG}(\cdot)$, $\text{AGG}(\cdot)$, and $\text{Read-Out}(\cdot)$ functions learn substantially different patterns for these two categories of actions. Specifically, taking the feature maps obtained by MSG(·) as a close look, the DGNN learns the largest intensity values for the joint of index 16 and 20, corresponding to the left foot and right foot, respectively, for the action category of walking; but the joint of index 8, corresponding to the left hand, for the action category of waving hands. This can be interpreted as the critical of the joints for recognizing these two actions, which is consistent with the human consciousness. In Fig. 4d, the categorical voting matrix of one round in 5-fold cross validation, corresponding to 95.8% video accuracy, is provided to visualize the classification results on all the test sub-sequences. The percentage of votes for the six actions in each test video is calculated, and the videos are re-ordered so that the diagonal blocks of the matrix represent the correct classification. The test result shows that only one video, indicated by the arrow, is misclassified.

## Discussions

**Scarce training labels.** We analyze the effectiveness of DGNN under the limited size of training labels, which is a common case in semi-supervised learning. With the same architecture settings, we compare the performance of DGNN with respect to the baselines under different sizes of training labels, including 1, 5, 10, 15, 20, and 25 labels per class. We plot the bar graph of test accuracy with error bars by performing 10 times evaluations for each size of training label in Fig. 5. The mean values of the results are shown in Supplementary Table S3. The DGNN architecture outperforms all baselines by a substantial amount for all label-scarce settings, especially at the small training-set size, e.g., only one label per class, which demonstrates the higher generalization ability with respect to other electronic computing approaches.

**Scalability of architecture.** The proposed DGNN architecture performs AGG(·) only once to directly consider high-order node features, which avoids the exponential neighborhood expansion issue in extracting long-range neighborhood information and facilitates the scalability for learning larger graphs. In principle, the head number of architectures can be scaled to arbitrary size, and the basic DPU modules, e.g., in Fig. 1c, can be horizontally stacked and interconnected with Y-couplers and strip waveguides to aggregate optical neural messages from the arbitrary size of neighborhoods. Moreover, the architecture has the flexibility to extend with multiple rounds of optical message passing by further stacking DPU modules. The DPU module can be scaled up by increasing the numbers of metaline layers and meta-atoms at each layer, and the input and output numbers of the DPU can be scaled up with additional optical modulators and waveguide crossings, e.g., in Supplementary Fig. S6. The accumulation of system error can be alleviated by re-training the output classifier, e.g., in Supplementary Fig. S8. Besides, the in-situ training approach[18] can also be applied to address the system errors by developing the on-chip DPU modules with programmable modulation coefficients, e.g., using the 1D indium tin oxide (ITO) for modulation[57].

**Computing density and energy efficiency.** It is worth noting that once the DGNN architecture design is optimized and fabricated physically, the on-chip optical devices for the computation of node and graph representations as well as the optical output classifier during the inference are all passive. Such the inference process for graph-based AI tasks is processed at the speed of light, limited only by the input data modulation and output detection rates, and consumes little energy compared with electronic GNNs. To be specific, assuming the DGNN transforms $n$-dimensional attributes into the $m$-dimensional optical neural messages for each node with $\text{MSG}(\cdot)$, aggregates optical features of $k$ nodes with $\text{AGG}(\cdot)$, and stacks $P$ heads for a $C$-class classification task. Therefore, the $\text{MSG}(\cdot)$ module of each node contains an $n \times m$ weight matrix for each node, the $\text{AGG}(\cdot)$ module in each head contains the sum of $k$ nodes of $m$-dimensional vectors, and the classifier contains an $mP \times C$ weight matrix. Therefore, each inference cycle of DGNN contains $(nmk + mk)P$ operations (OPs) for feature extractions and $mPC$ operations for the classification, i.e., having the total operations of $(nk + k + C)mP$. Considering a 100 GHz data modulation and detection rate[22], the computing speed of DGNN is $(nk + k + C)mP \cdot 10^{11}$ OPs/s. Assuming the typical light source power of 10 mW, the energy efficiency of DGNN is $(nk + k + C)mP \cdot 10^{13}$ OPs/J. For the node classification settings with $n = 20, m = 2, k = 8, P = 4, C = 8$, the computing speed is $140.8 \text{ TOPs}^{-1}$ and energy efficiency is $14.080 \text{ POPs}^{-1}\text{W}^{-1}$. For the DPU module in Fig. 2 with a size of $72.85 \mu m \times 27 \mu m$, which performs the $\text{MSG}(\cdot)$ function with $3 \times 2$ weight matrix, the computing density is $305 \text{ TOPs}^{-1}\text{mm}^{-2}$. Notice that the energy efficiency and computing density of the state-of-the-art GPU Tesla V100 are $100 \text{ GOPs}^{-1}\text{W}^{-1}$ and $37 \text{ GOPs}^{-1}\text{mm}^{-2}$, respectively[58]. The proposed approaches can achieve orders of magnitude improvements.

**Limitations.** In this study, the optical feature aggregation in DGNN is realized by using the $2 \times 1$ optical Y-coupler with a combining ratio of $50:50$, which does not support the assigning of different weights for different neighboring nodes, i.e., the weighted sum. Although the average feature aggregation has already achieved remarkable performance in both node- and graph-level classification tasks, we could utilize the on-chip amplitude modulator, e.g., phase changing materials[25], to further increase the flexibility and model capacity of optical feature aggregation in the future. Another limitation is that the proposed DGNN architecture uses a linear model for optical message passing. Although existing works have demonstrated the possibility of implementing the optical nonlinear activation functions[59-61], the nonlinear operation is not critical in GNNs as studied in the previous work[54]. This can be proved by the remarkable model performance that DGNN has already achieved in the real-world benchmark datasets. For example, DGNN almost achieves the state-of-the-art performance on Amazon Photo under large scarce training labels, and significantly outperforms the electronic GNNs under the scarce label settings. Therefore, including nonlinear activation function in DGNN is left for future work as the potential to further enhance the model learning capability.

## Conclusion

In summary, we take the first step to present the optical deep learning architecture, i.e., DGNN, that can perform the all-optical graph representation learning over the graph-structured data for the high-accurate node- and graph-level classification tasks. The architecture is designed by utilizing the integrated DPU for extracting optical neural messages of graph nodes and on-chip optical devices for passing and aggregating the messages. We verify the functionalities of DGNN with both the analytical and FDTD evaluations. The results demonstrate the comparable and even superior classification performance than the electronic GNN and achieve orders of magnitude improvement on computing performance than the electronic computing platform. We expect that our work will inspire the future development of advanced optical deep learning architectures with integrated photonic circuits beyond the Euclidean domain for high-efficient graph representation learning.

# Methods

**PPRGo model**. PPRGo[38] implements the $\text{MSG}(\cdot)$ with a neural network to transform node attributes, and performs a single round $\text{AGG}(\cdot)$ for each target node to directly aggregate information from the top-$k$ neighboring nodes ordered by the nodes' personalized PageRank score. The personalized PageRank matrix is analytically defined as: $\mathbf{\Pi} = \alpha\big(\mathbf{I} - (1-\alpha)\tilde{\mathbf{A}}\big)^{-1}$, where $\alpha \in (0,1]$ is the teleport probability of personalized PageRank, $\tilde{\mathbf{A}} = (\mathbf{D}+\mathbf{I})^{-1/2}(\mathbf{A}+\mathbf{I})(\mathbf{D}+\mathbf{I})^{-1/2}$ is the symmetric normalized adjacency matrix with added self-loops, $\mathbf{A}$ denotes the adjacency matrix, $\mathbf{D}$ denotes the degree matrix, and $\mathbf{I}$ denotes the identity matrix. The $i$-th row of $\mathbf{\Pi}$, denoted by $\mathbf{\Pi}_i$, is the personalized PageRank scores of all the graph nodes with respect to node $v_i$. PPRGo performs the $\text{AGG}(\cdot)$ for node $v_i$ by only summing the features of nodes that are the top-$k$ largest entries in $\mathbf{\Pi}_i$, where the aggregated node features are fed to the output classifier to predict the label of node $v_i$. Note that the calculation of $\mathbf{\Pi}_i$ is a procedure of data pre-processing, which only needs to be calculated once and can be implemented with fast algorithms[38]. We applied two variants of PPRGo, i.e., the PPRGo-S and PPRGo-WS (see Table 1). PPRGo-S uses the aggregator that directly sums up the neighboring features. In contrast, PPRGo-WS uses the aggregator that performs the weighted sum of neighboring features according to the personalized PageRank scores.

**Generating the synthetic graphs.** The stochastic block model (SBM)[53] is a widely used generative graph model in network analysis. We evaluated the effectiveness of our all-optical graph representation learning by generating the synthetic SBM graph with 300 nodes to reduce the computational complexity and the requirement of computing resources during the architecture evaluation using FDTD. Specifically, the 300 nodes were assigned to 3 communities (categories), and node attributes of each community were generated from the corresponding 3-dimensional multivariate Gaussian distribution. The simplest SBM has two parameters $p$ and $q$, corresponding to intra-class link probability and inter-class link probability, respectively, with the graph generation rule as follows:

$$a_{ij}|y_i, y_j \sim \begin{cases} \text{Bernoulli}(p), & \text{if } y_i = y_j \\ \text{Bernoulli}(q), & \text{if } y_i \neq y_j \end{cases}, \qquad (4)$$

where $y_i$ and $y_j$ denote the category of nodes $v_i$ and $v_j$. In this work, we set $p = 0.1$ and $q = 0.005$. We randomly selected 5 labeled nodes per category for training, and the left 285 nodes were used for the test. The generated graph is illustrated in Fig. 2a.

**Pre-processing of benchmark graphs for node classification**. Cora-ML and Citeseer are the document co-citation networks, in which each node represents a document and edges are citations between them. Amazon Photo is a segment of the Amazon Co-purchase graph[52], where nodes represent goods and edges denote that the two goods are frequently bought together. All three graphs have node attributes encoded by bag-of-words. Following previous works for node classification on benchmark graphs[62], we randomly selected 1000 nodes as the test set for each benchmark dataset with the remaining nodes for training. To reduce the number of input strip waveguides for the DPU, we adopted the principal component analysis (PCA) to pre-process the node attributes and reduce their dimensions. We set the dimension of the node attribute to be 20, and the values were scaled to be compatible with the optical system, i.e., encoding the node attributes into either the amplitude (rescale to $[0, 1]$) or phase (rescale to $[0, 2\pi]$) of coherent optical waves. The overview of dataset statistics is summarized in Supplementary Table S1.

**DPU settings.** The integrated DPU utilizes the successive layers of diffractive metalines to modulate the optical wavefront. Each metaline comprises diffractive meta-atoms, i.e., the array of rectangle silica slots etched in the silicon membrane of SOI substrate, as shown in Supplementary Fig. S1a. The height and width of a slot determine the phase and amplitude modulation coefficients of a diffractive meta-atom. We adopted the Rayleigh-Sommerfeld diffraction for analytical modeling the optical wave propagation and modulation[14], and the FDTD evaluations were performed via Lumerical FDTD software (Lumerical Inc.). The working wavelength of our architecture was set to be 1.55 μm. To facilitate the training of DPU and improve the modulation accuracy, we used the subwavelength height for the silica slot and fixed it to be 400 nm, with which the silica slot width was optimized within [0, 100] nm under the fix slot period of 300 nm, corresponding to the optimizing of the phase modulation range of [0, 1.55] rad (see Fig. S1b and S1c). Moreover, considering the fabrication capability of existing silicon photonics foundry and to reduce the modeling deviation, we also adopted the binary modulation that the width of the slot was quantized to take value from $\{0, 100\}$ nm and set the slot width of every 3 consecutive meta-atoms to be the same. The input and output planes were divided into regular intervals with the numbers equivalent to the numbers of input and output waveguides, where each waveguide was placed at the central position of each interval.

In the task of node classification, the DPU module was set to have 3 and 4 layers of metalines with a layer distance of 20 μm and 100 μm, respectively, to perform the feature transformation for the synthetic and benchmark databases, respectively; each metaline was to comprise 90 and 600 meta-atoms, respectively, corresponding to the metaline length of 27 μm and 180 μm, respectively. The input and output planes were coupled with 3 waveguides and 2 waveguides, respectively, for the synthetic database; and 20 waveguides and 2 or 8 waveguides, respectively, for the benchmark database. Besides, for the benchmark database, the output classifier DPU module of DGNN-O architecture was set to have 6 layers of metalines with other settings the same as the feature transformation DPU module. In the task of skeleton-based human action recognition, the feature transformation DPU module was set to have 6 layers of metalines and 3 input optical waveguides for encoding 3D joint coordinates with other settings the same as the node classification on real-world graphs. Similarly, the read-out function of each head was implemented with the DPU module with 5 layers of metalines.

**Training details of DGNN.** All DGNN models were numerically implemented and trained based on Python (v3.6.8) and TensorFlow (v1.12.0, Google), and the PCA was implemented using Scikit-learn (v 0.23.2). The diffractive wave propagation was analytically modeled with the Rayleigh-Sommerfeld diffraction model implemented using the angular spectrum method[14]. The modulation coefficients of diffractive layers were optimized during the training, where the phase modulation coefficient of each diffractive meta-atom was correlated with amplitude modulation coefficients and determined by the slot width (see Supplementary Fig. S1c). We adopted the Adam optimizer[63] to perform the gradient descent and error backpropagation. The loss function of DGNN-E was the softmax cross-entropy between the electronic output and the one-hot ground-truth labels, while the loss function of DGNN-O was the mean-squared-error (MSE) between the detected intensity values on the output plane and the target, i.e., 1 for the position of the target detection region and 0 for the others regions). The learning rate was set to 0,1, 0.01, and 0.005 for the node classification with DGNN-O, node classification with DGNN-E, and skeleton-based human action recognition, respectively. The re-training procedure used a learning rate of 0.1. We used the full-batch training fashion in the node classification, while the batch size was set to 32 in the task of skeleton-based action recognition. Besides, for training DGNN-E with binary modulation, the modulation coefficient of each meta-atom was computed with an extra rounding operation.

**Training details of electronic models.** Similarly, all implementations of electronic models were based on Python (v3.6.8), TensorFlow (v1.12.0, Google), and Scikit-learn (v 0.23.2). The PCA classification results were obtained by using a linear classifier to the pre-processed node attributes. The MLPs were configured by using a hidden layer with a size of 8 and activated by the ReLU nonlinear activation function. The electronic PPRGo GNNs used the MLP for implementing the feature transformation. All the electronic models used the softmax cross-entropy between predictions and targets as the loss function. The learnable parameters were optimized by using an Adam optimizer with a learning rate of 0.01 and a training epoch of 10000. To avoid the over-fitting, we applied the $L_2$-norm regularization on the learnable weights with weight decay tuned in the search space {0.0001, 0.0005, 0.001, 0.005}.

**Analytical modeling of DPU with system errors.** To reduce the modulation deviations between the analytical model and FDTD due to the un-continuous change of parameters between adjacent meta-atoms, every 3 consecutive meta-atoms in the metalines was restricted to be the same (see Supplementary Fig. S2). Other error sources that cause the model deviations include the mutual coupling between adjacent meta-atoms, the reflection between metalines, and the fabrication errors during the semiconductor process. During the evaluation, we modeled the system error by including the Gaussian noise with a standard deviation of 0.3 to the trained phase modulation coefficients and the amplitude modulation coefficients. Moreover, the architecture still performs well even under more significant Gaussian noise, demonstrating its robustness to the system errors (see Supplementary Fig. S8).

**Transductive learning and inductive learning.** In the task of semi-supervised node classification, which is also termed transductive learning, all the nodes and the graph structure are available. While in supervised learning, i.e., inductive learning, all the test node are unavailable. For the inductive learning in this work, all 1000 test nodes, including their features and graph structures, are unseen during the training. In other words, we delete all test nodes with their associated edges to obtain the training set. During the test, we recover the original graph to perform the inference.

**Acknowledgements**

This work is supported by the National Natural Science Foundation of China (No. 62088102), the National Key Research and Development Program of China (No. 2020AAA0130000), the Beijing Municipal Science and Technology Commission (No. Z181100003118014), and the Tsinghua University Initiative Scientific Research Program.

**Author contributions**

Q.D., X.L. and H.X. initiated and supervised the project. X. L., R.Y. and T.Y. conceived and designed the research. T.Y. and R.Y. implemented the algorithm and conducted the numerical experiments. R.Y., T.Y. and Z.Z. processed the data. X.L., R.Y., T.Y. and Z.Z. analyzed and interpreted the results. All authors prepared the manuscript and discussed the research.

**Additional information**

Correspondence and requests for materials should be addressed to X.L. (lin-x@tsinghua.edu.cn), H.X. (xionghongkai@sjtu.edu.cn) or Q.D. (daiqh@tsinghua.edu.cn).

Supplementary information is available for this paper, including:

Figures S1 to S12

Tables S1 to S3

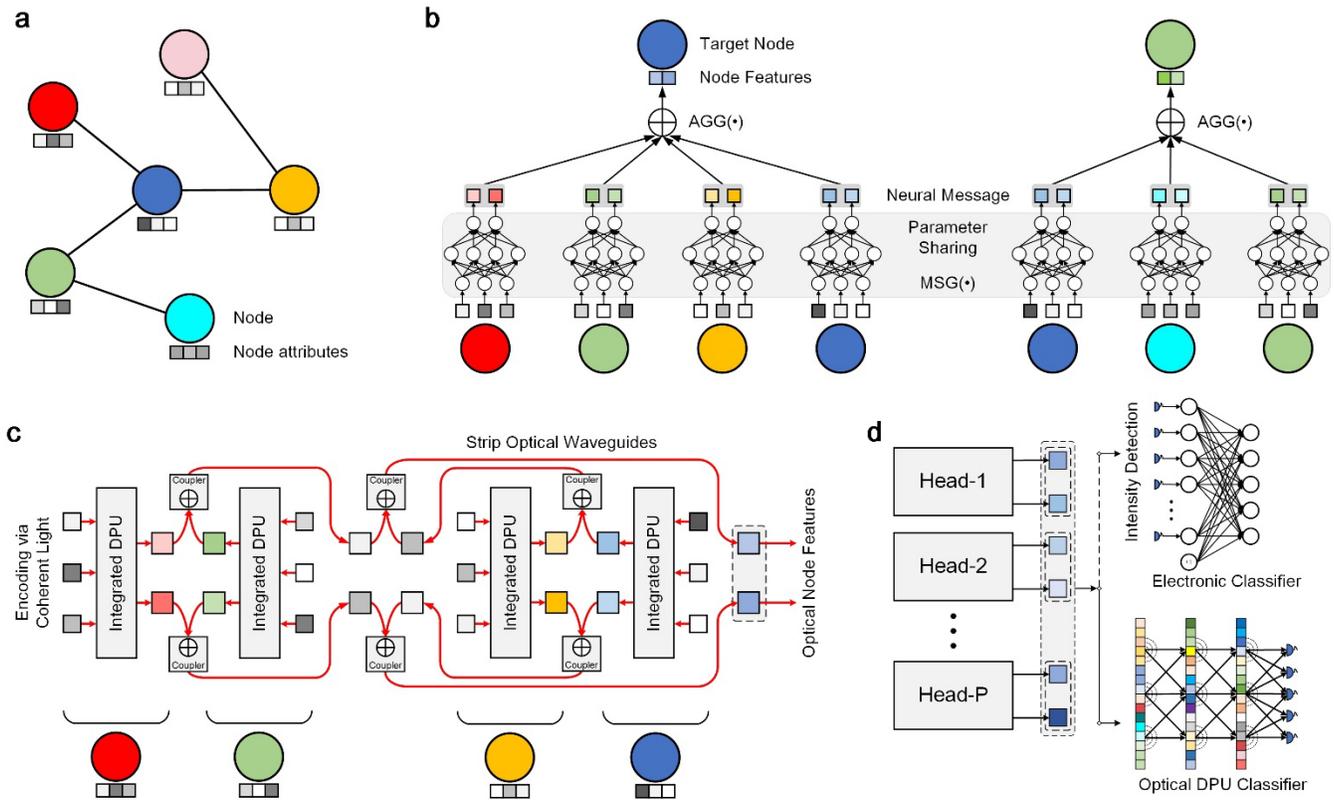

**Figure 1. The architecture of optical DGNN.** (a) An exemplar graph with six nodes and five edges. Each node has three-dimensional attributes. (b) The schematic illustration of a single round message passing of the GNN for target graph nodes, including the feature transformation and aggregation. (c) An all-optical architecture illustration for graph representation learning, where node features are encoded into amplitude or phase of the light in optical waveguides and transformed by the integrated DPUs. The transformed optical node features are aggregated using the optical couplers. The architecture is scalable for large graphs with a large number of nodes. (d) A multi-head strategy is adopted to extract high dimensional optical node features, based on which the node and graph classification tasks are performed using either electronic or optical DPU classifier, resulting in the DGNN-E or DGNN-O architecture. Each head is a structure like (c) that produces two-dimensional optical features.

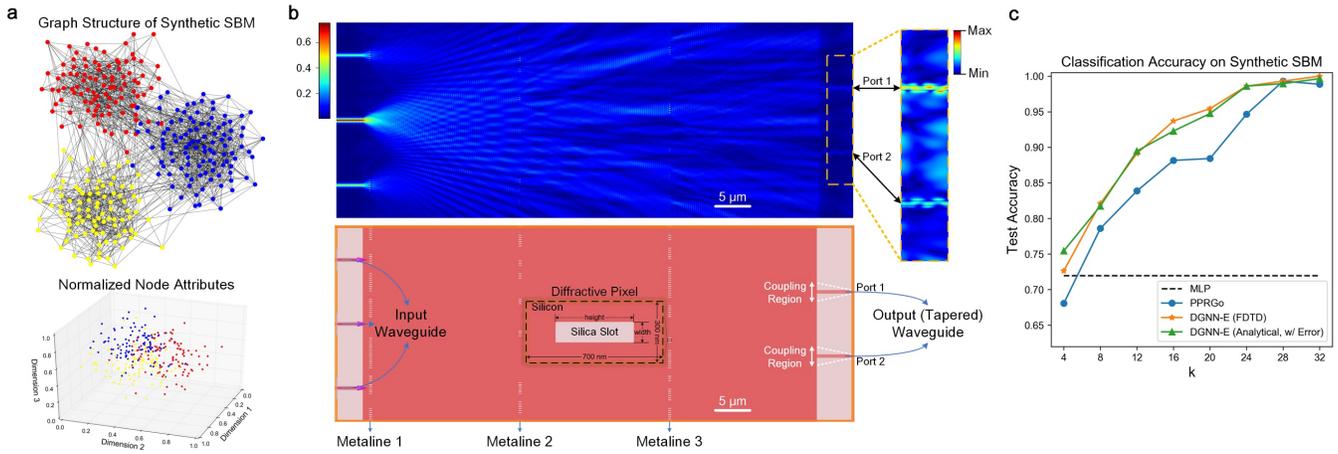

**Figure 2. Semi-supervised node classification on a synthetic graph.** (a) Top: The synthetic SBM graph with 300 nodes, where different colors denote different communities, i.e., categories. Bottom: The three-dimensional node attributes of different communities are generated from different multivariate Gaussian distributions and normalized to $[0, 1]$. (b) Top: The electric field of the DPU evaluated with FDTD that performs the $\text{MSG}(\cdot)$ function on a target node of the synthetic SBM (see Supplementary Fig. S5 for the tapered output waveguide implementation that significantly improves the DPU power transmission rate). Bottom: The corresponding DPU module is implemented in the silicon membrane of the SOI chip. (c) Classification accuracy of DGNN-E under different top-$k$ nodes for feature aggregation with comparisons to MLP and PPRGo electronic models.

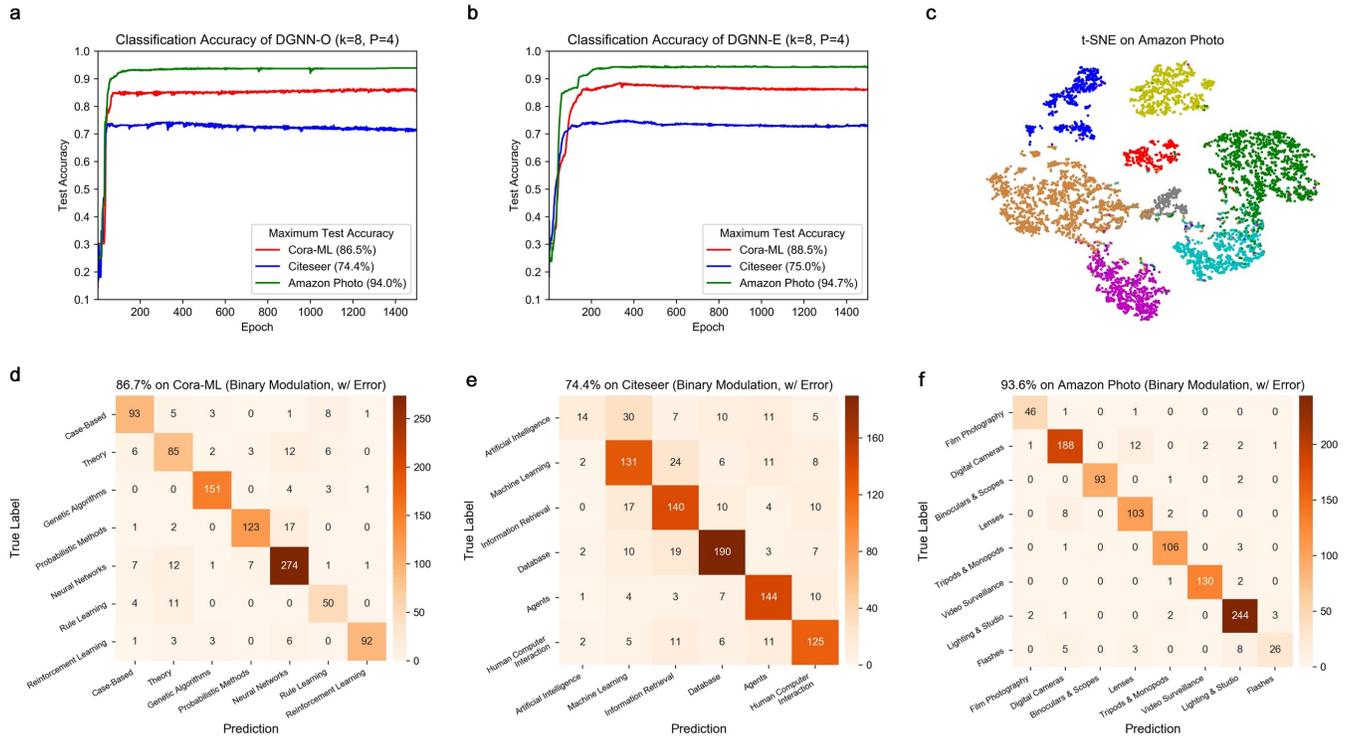

**Figure 3. Semi-supervised node classification on three benchmark graph databases.** (a) Test accuracy convergence plots of DGNN-O with optical DPU output classifier. (b) Test accuracy convergence plots of DGNN-E with electronic output classifier. (c) t-SNE visualization of node representations of DGNN-E on the Amazon Photo dataset. (d, e, f) Confusion matrices of DGNN-E classification result on three graphs with binary modulation and system errors by including Gaussian noise with a standard deviation of 0.3.

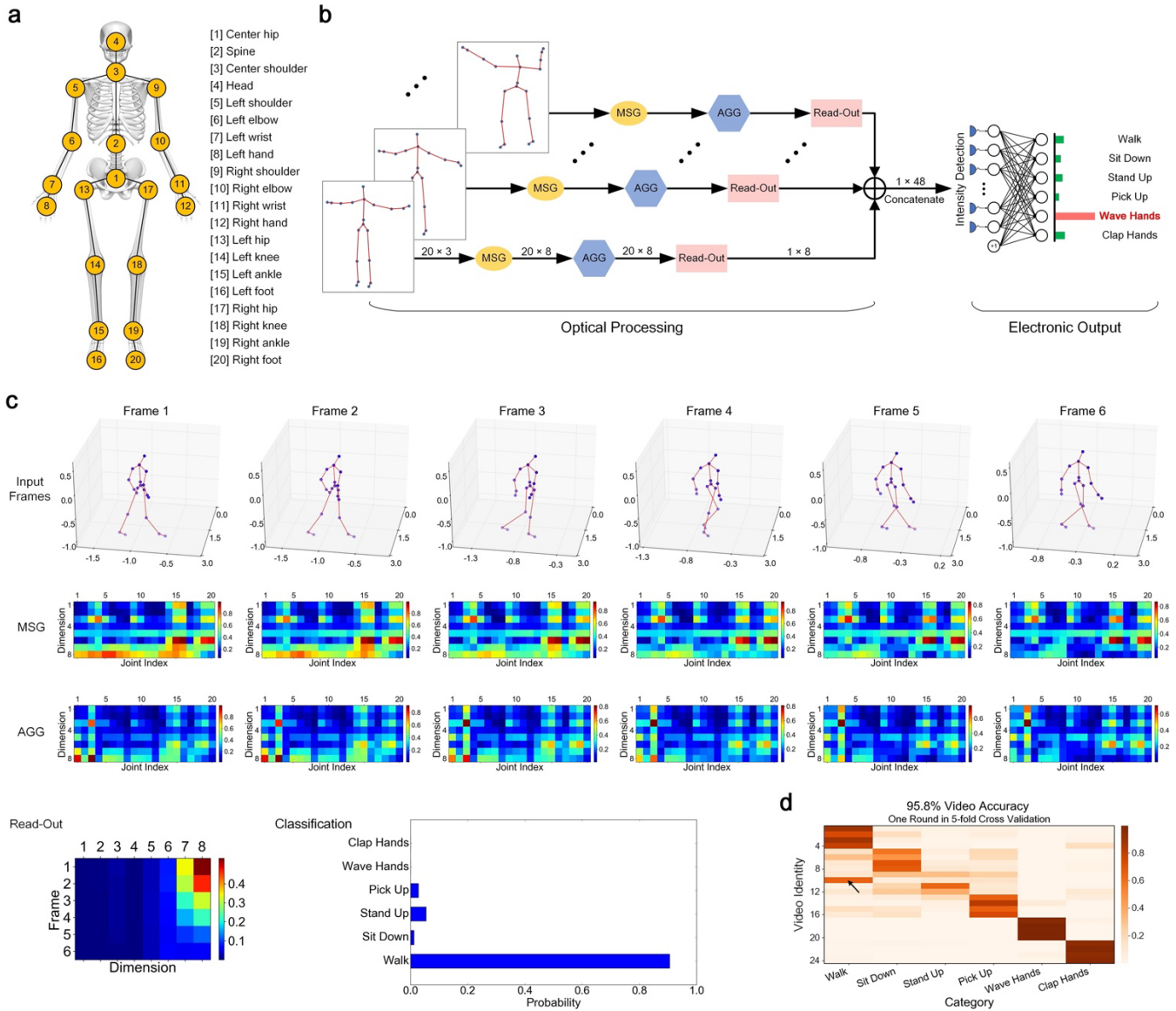

**Figure 4. Graph classification of the DGNN on the task of action recognition.** (a) Graph structure of skeleton data captured by Kinect V1. (b) The schematic of DGNN architecture for skeleton-based human action recognition. (c) Visualizing results of a selected sub-sequence from the test set for performing the action category of the walk. The normalized amplitude of each frame processed after optical MSG(·), AGG(·), the $L_2$-normalized intensity values after optical Read-Out(·), and the classification result are shown. (d) The inference results of all the test sub-sequences in one round of 5-fold cross-validation. The arrow indicated slot is the only misclassified video of the database.

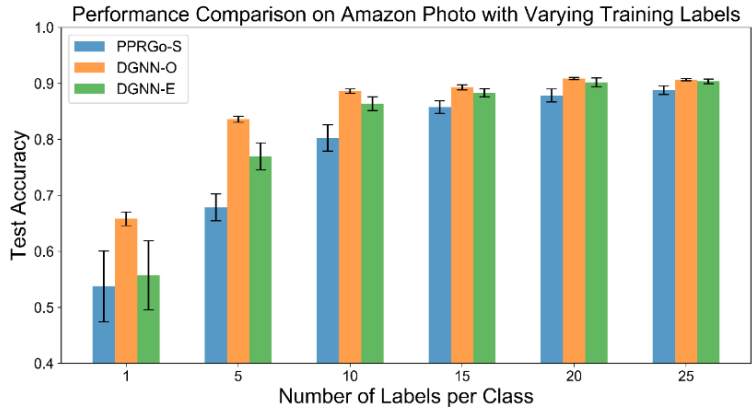

**Figure 5. Classification on Amazon Photo with scarce training labels.** Performance of architecture under different training-set sizes are evaluated. Both the DGNN-O and DGNN-E architectures consistently outperform the counterpart electronic PPRGo-S GNN model and demonstrate superior robustness and generalization ability.

**Table 1. Semi-supervised node classification results on three benchmark graphs.** The classification accuracy (%) of DGNN architecture are obtained by setting $k=8$ and $P=4$.

| Dataset | Cora-ML | Citeseer | Amazon Photo |
|---|---|---|---|
| **PCA** | 79.4 | 70.7 | 90.8 |
| **MLP** | 78.8 | 70.9 | 91.2 |
| **PPRGo-S** | 86.9 | 74.1 | 94.9 |
| **PPRGo-WS** | 87.6 | 75.5 | 94.9 |
| **DGNN-O** | 86.5 | 74.4 | 94.0 |
| **DGNN-E** | 88.5 | 75.0 | 94.7 |
| **DGNN-E (Binary Modulation, w/ Error)** | 86.7 | 74.4 | 93.6 |